\documentclass[12pt]{article}
\usepackage{amsmath,amssymb,amsthm,color}
\usepackage{graphicx}
\usepackage{cite}
\usepackage{epsfig}
\renewcommand{\baselinestretch}{1.66}
\begin{document}
%
\title{Magneto-optical selection rules of curved graphene nanoribbons and carbon nanotubes \\}
\author{
\small Chiun-Yan Lin$^{a}$, Jhao-Ying Wu$^{a,\dag}$, Cheng-Pong Chang$^{b,~\ddag}$, Ming-Fa Lin$^{a,*}$ $$\\
\small  $^a$Department of Physics, National Cheng Kung University, Tainan, Taiwan 701\\
\small  $^b$Center for General Education, Tainan University of Technology, Tainan, Taiwan 710 \\
 }
\renewcommand{\baselinestretch}{1.66}
\maketitle

\renewcommand{\baselinestretch}{1.66}
\begin{abstract}
We derive the generalized magneto-absorption spectra for curved graphene nanoribbons and carbon nanotubes by using the Peierls tight-binding model. The main spectral characteristics and the optical selection rules result from the cooperative or competitive relationships between the geometric structure and a magnetic field. In curved ribbons, the dominant selection rule remains unchanged during the variation of the curvature. When the arc angle increases, the prominent peaks are split, with some even vanishing as the angle exceeds a critical value. In carbon nanotubes, the angular-momentum coupling induces extra selection rules, of which more are revealed due to the increase of either (both) of the factors:  tube diameter and field strength. Particularly once the two factors exceed certain critical values, the optical spectra could reflect the quasi-Landau-level structures. The identifying features of the spectra provide insight into optical excitations for curved systems with either open or closed boundary condition.
\vskip 1.0 truecm
\par\noindent

\par\noindent \dag Corresponding author. {~ Tel:~ +886-6-275-7575;~ Fax:~+886-6-74-7995.}\\~{{\it E-mail address}: l2894110@mail.ncku.edu.tw (J.Y. Wu)}

\par\noindent ~\ddag Corresponding author. {~ Tel:~ +886-6-253-2106;~ Fax:~+886-6-254-0702.}\\~{{\it E-mail address}: t00252@mail.tut.edu.tw (C.P. Chang)}

\par\noindent * Corresponding author. {~ Tel:~ +886-6-275-7575;~ Fax:~+886-6-74-7995.}\\~{{\it E-mail address}: mflin@mail.ncku.edu.tw (M.F. Lin)}

\end{abstract}

\pagebreak
\renewcommand{\baselinestretch}{2}
\newpage

{\bf 1. Introduction}
\vskip 0.3 truecm

Graphene is a one-atom-thick hexagonal lattice sheet made up of carbon atoms [1].
It is a flexible membrane that can be curled or twisted without loosing its unique electronic and atomic structural properties. To date, many studies have focused on a host of curled and folded structures of graphene-related materials, such as fullerenes [2], graphene bubbles [3], graphene ripples [4,5], carbon nanotubes [6-13], carbon tori [14-17], carbon nanoscrolls [18], coiled carbon nanotubes [19], and deformed graphenes [20-23]. These materials have specific geometric symmetries and dimensions with either open or closed boundaries. Of interest are the physical properties resulting from the cooperative or  competitive relationship between the geometry and a magnetic field---properties such as the optical selection rule [9-11], magneto-transport [21,22] and the quantum Hall effect [18]. Recently, a curved graphene nanoribbon, an intermediate system between a carbon nanotube and a graphene nanoribbon, has been fabricated from carbon nanotubes by using chemical oxidation [24-26], argon plasma etching [27], mechanical cutting [28], and metal particle-assisted [29] and electrical current-induced [30] unzipping procedures. Curving a graphene ribbon provides an ideal opportunity to study the effects of a non-uniform magnetic field on a structure with an open boundary condition [31]. To illustrate the optical response under the influence of the geometry and the magnetic field, and the interplay between open and closed boundary conditions, we thoroughly investigate the magneto-optical spectra of curved graphene nanoribbons and carbon nanotubes. This study serves as a first step toward fully understanding the magneto-electronic and magneto-optical properties of curved carbon nanostructures.

The 1D electronic properties of graphene nanoribbons strongly depend on both quantum confinement and edge-structure effects [32-37]. In the presence of a sufficiently large perpendicular magnetic field, the 1D energy bands evolve into dispersionless quasi-Landau levels (QLLs), regardless of the edge structure [38-40]. Unlike in a monolayer graphene [1,41], the formation of the QLLs in graphene nanoribbons is limited to lower energies. Studies on absorption spectra have been performed to demonstrate that the prominent peaks result from the inter-QLL transitions [42]. As a flat ribbon is bent into a curved one, the QLLs are depressed due to the weakened magnetic quantization [31]. They are entirely changed into oscillating parabolic bands, of which the asymmetric energy dispersions can be attributed to a non-uniform effective magnetic field. It is obvious that the number of van Hove singularities (vHs) in the density of states (DOS) is increased, and thus the available optical excitations related to the various vHs are expected to induce rich and complex optical spectra. Since spatial changes are reflected in the wavefunctions, it is worthwhile validating whether the specific selection rule for inter-QLL transitions, $\Delta n=n^{c}-n^{v}=\pm1$ (with $n^{c,v}$ being the Landau level index), survives or whether other selection rules appear under certain circumstances through the effects of the magnetic field and the geometric structure.

Since a carbon nanotube has a closed periodic structure, the essential difference between a tube and a curved ribbon arises from the distinct boundary conditions. The cylindrical symmetry leads to the quantization of the angular momentum $J$ [43-45]. When the electric polarization is parallel to the tube axis, the allowed interband transitions are those between the occupied and unoccupied states with the same $J$ [46-52]. In an axial magnetic field, the peak splitting and the periodicity of Aharonov-Bohn oscillations theoretically predicted in interband magneto-absorption spectra [8-10] were experimentally verified [11], whereas the selection rule remains unchanged. On the other hand, a transverse magnetic field induces a coupling of the independent angular-momentum states with one another. This might trigger available excitations between various valence and conduction bands. The optical spectra and the selection rules in cases of different field strengths and tube radii are worthwhile exploring further. Recently, experimental observations on the QLL states of carbon nanotubes were achieved via magneto-transport measurements in a carbon-nanotube-based Fabry-Perot resonator under a very strong field [53,54]. Also crucial is the issue of the optical response in this case of extremely strong coupling.

The paper is a comprehensive study that elaborates on the magneto-absorption spectra and the optical selection rules during the zipping process that turns graphene nanoribbons into carbon nanotubes. The purpose is to provide an ideal means for exploring the magneto-optical response of the curved graphene system through the transformation of an open boundary into a closed boundary system. The influences of varying the geometric curvature and applying a magnetic field on the spectra of curved nanoribbons are discussed first. It is shown that the prominent peaks of $\Delta n=\pm1$ start to split and become separated from one another as the geometric curvature becomes larger. Moreover, we explore the correlation between the prominent peak structures (optical selection rules) of carbon nanotubes and the degree of angular-momentum coupling, i.e., extra peaks and selection rules corresponding to the coupling. Their dependence on field strength and ribbon width (tube diameter) is investigated in greater detail below. We discuss how the spectra of curved ribbons can be partly obtained from an interpolation between flat ribbons and cylindrical nanotubes. The resulting extraordinary properties can be explained by the complex relationship among the aforementioned effects.

This paper is organized as follows. In section 2, the Hamiltonian is built from the Peierls tight-binding model in the presence of a perpendicular magnetic field. The spectral absorption function is evaluated within the gradient approximation. Discussions on the magneto-electronic properties and the electronic wavefunctions are offered in section 3.1. In section 3.2, the curvature-dependent magneto-optical spectra for different field strengths are studied. Meanwhile, the dependences of the absorption frequency on the field strength and the geometric curvature are also investigated. Finally, section 4 contains concluding remarks.

\vskip 0.6 truecm
\par\noindent
{\bf 2. Methods }
\vskip 0.3 truecm

Carbon-related nanostructures in both the ribbon and tube shapes can be categorized into chiral or achiral configurations, depending on the cut and wrap angle [44,45]. Most of the studies have mainly focused on the highly symmetric achiral zigzag and armchair forms [12-14,16,17,31]. Chosen as a study model in this work, a zigzag nanoribbon is a zigzag-edge graphene strip with a nanometer-scale width. The bond length between carbon atoms is $b$=1.42 {\AA }. The periodic length along the $\hat{y}$ direction is $I_{y}=\sqrt{3}b$, and the first Brillouin zone is defined as $-1\leq k_{y}\leq 1 $ in the unit of $\pi /I_{y}$. The number of the zigzag lines $N$ determines not only the ribbon width $W$($=(3N-2)b/2$), but also the number of carbon atoms in the primitive unit cell ($N$ A atoms and $N$ B atoms). As shown in Fig. 1(a), a curved graphene nanoribbon is characterized by the bending arc angle $\theta $ and the radius $R(=W/\theta )$, where $\theta $ represents the curvature of the ribbon.
\begin{figure}[htbp]
\center
\rotatebox{0} {\includegraphics[width=14cm]{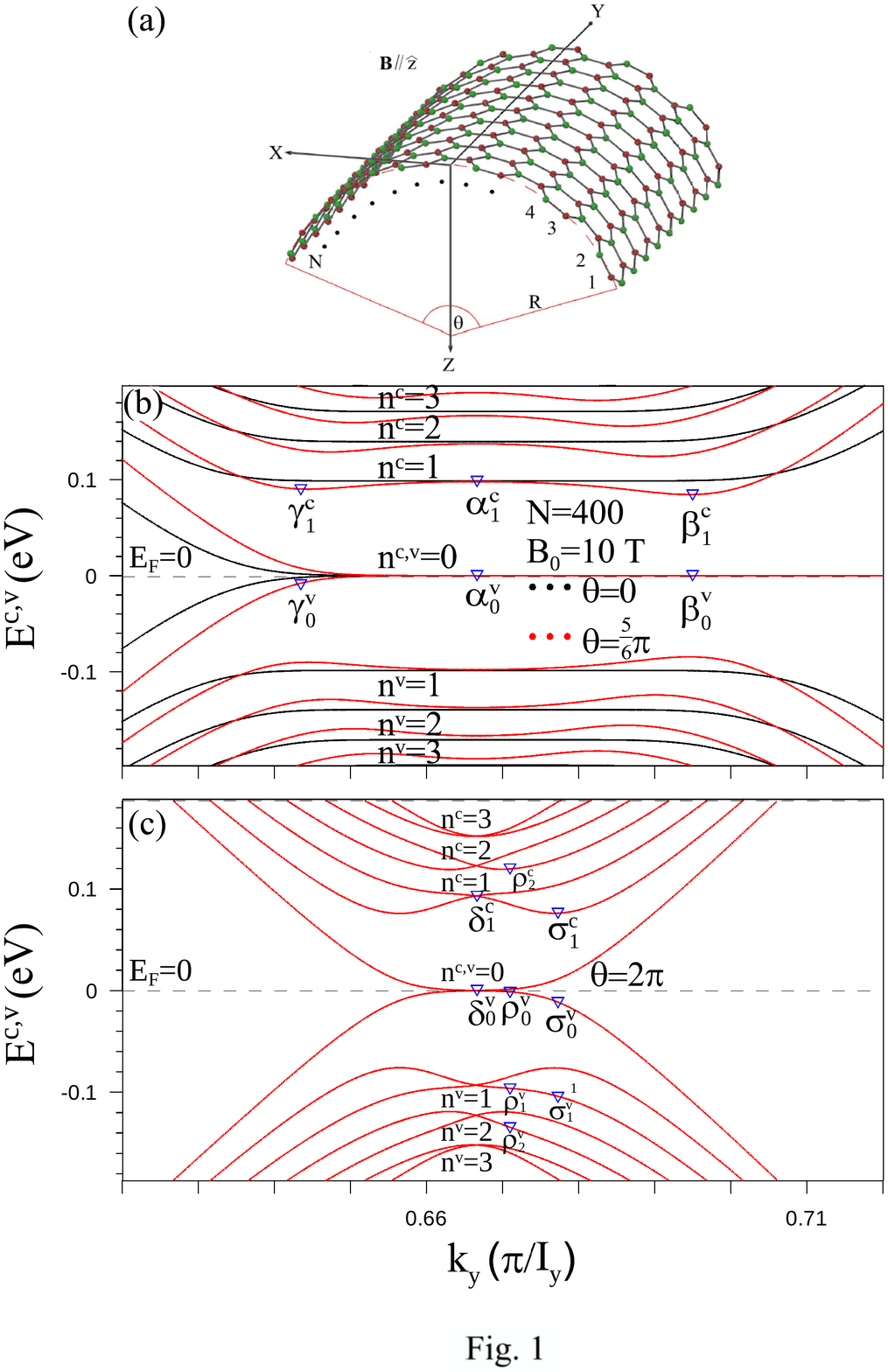}}
\caption{(a) A curved graphene nanoribbon, with the curvature radius $R$ and the bending arc angle $\theta$, in the uniform magnetic field $\textbf{B}=B_{0}\hat{z}$, perpendicular to the plane tangent to the ribbon bottom. At $B_{0}=10$ T, the band structures of (b) the $N=400$ curved zigzag ribbon with $\theta $=$5\pi/6$ and (c) the (200, 200) armchair carbon nanotube are plotted. Also shown in (b) is that of the flat zigzag ribbon for a comparison.}
\label{Figure 1}
\end{figure}
The low-energy electronic structures mainly arising from the $2p_{z}$ orbitals are calculated by the nearest-neighbor tight-binding model with the curvature effect.
The Bloch wavefunction is expressed as
\begin{equation}
\left\vert \Psi_{k_{y}}\right\rangle = \sum^{N}_{m=1} A_{m} \left\vert\
a_{mk_{y}}\right\rangle+ \sum^{N}_{m=1}B_{m} \left\vert\
b_{mk_{y}}\right\rangle,
\end{equation}%
where $\left\vert \ a_{mk} \right\rangle$ and $\left\vert \
b_{mk}\right\rangle$ for m=1,2,3,.....,$N$ are the tight-binding functions
related to the periodic $A$ and $B$ atoms. Due to the misorientation of the $2p_{z}$ orbitals on the cylindrical surface, the curvature modifies the hopping integral between two atoms. The hopping integral is given by $\gamma _{\alpha}$($\theta _{\alpha}$)=$V_{pp\pi }$ $\cos $($\theta _{\alpha}$)+$4(V_{pp\pi}-V_{pp\sigma}$) [${R}/{b}$ $\sin ^{2}$(${\theta_{\alpha}}/{2}$)]$^{2}$, where $\alpha$(=1, 2, 3) corresponds to the three nearest neighbors, and $\theta _{\alpha}$ ($\theta _{1}={b}/{R}$; $\theta _{2}=\theta_{3}={b}/{2R}$) represents the arc angle between two nearest-neighbor
atoms. The Slater-Koster parameters $V_{pp\pi}$$(=-2.66$ eV) and $%
V_{pp\sigma}$$(=6.38$ eV) are, respectively, the hopping integrals of the $\pi$ and $\sigma$ bonds between $2p$ orbitals in a flat plane [55]. The curvature effects taken into account in this model can explain the low-energy electronic properties of carbon nanotubes [9,11,43,44]. For the $\theta=0$ case, the hopping integral becomes $V_{pp\pi} $$=-2.66$ eV since $2p_{z}$ orbitals exclusively contribute to the $\pi$ components.
When a uniform magnetic field along the z-axis is applied to a curved zigzag ribbon, an extra Peierls phase factor $e^{i\Delta G_{\mathbf{R}}}$ is introduced in the Hamiltonian matrix element between sites $\mathbf{R_{i}}$ and $\mathbf{R_{j}}$, where $\Delta G_{\mathbf{R}}=$ (${{2\pi}/{{\phi }_{0}}}$) $\int_{\mathbf{R_{i}}}^{\mathbf{R_{j}}}{\mathbf{A}}\cdot d{\mathbf{l}}$ and the flux quantum $\phi _{0}=hc/e$ [56]. The vector potential in terms of the cylindrical coordinates $(r,\Phi ,y)$ is ${\mathbf{A}}\mathbf{=}B_{0}r\sin (\Phi )$ $\widehat{y}$. ${\mathbf{A}}$ is independent of $y$ so that $k_{y}$ remains a good quantum number. The $2N\times 2N$ Hamiltonian matrix built from the subspace spanned by the $2N$ tight-binding functions
\{$\left\vert\ a_{1k_{y}}\right\rangle,\left\vert\
b_{1k_{y}}\right\rangle,\left\vert\ a_{2k_{y}}\right\rangle,\left\vert\
b_{2k_{y}}\right\rangle,.......\left\vert\
a_{Nk_{y}}\right\rangle,\left\vert\ b_{Nk_{y}}\right\rangle$\} is
\begin{eqnarray}
H_{ij}=\left\{
\begin{array}{cc}
2\gamma _{2}\cos \{k_{y}\frac{\sqrt{3}b}{2}+\Delta G_{\mathbf{R}}\} &  \\
\gamma _{1} &  \\
0 &
\end{array}%
\begin{array}{c}
{\textmd{for}} \\
{\textmd{for}} \\
{\textmd{others}.}%
\end{array}%
\begin{array}{ccc}
{\textmd{j=i+1},} & {\textmd{j}} & {\textmd{is even} ,} \\
{\textmd{j=i+1},} & {\textmd{j}} & {\textmd{is odd} ;} \\
&  &
\end{array}%
\right\}.
\end{eqnarray}
$\Delta G_{\mathbf{R}}={2\sqrt{3}\pi bR^{2}}$$\{\cos
[3b(i-1/3)/4R-\theta /2]-cos[3b(i-1)/4R-\theta /2]\}$.
In the case of $\theta =2\pi$, i.e., an armchair carbon nanotube, the periodic boundary condition due to the cylindrical symmetry is imposed in the azimuthal direction. The tube diameter is determined by the number of zigzag lines in the cross section, and the matrix element $H_{i=1j=2N}=$ $\gamma _{1}$ is added in Eq. (2). The state energy $E^{c,v}(k_{y},n)$ and the wavefunction $\psi
^{c,v}(k_{y},n)$ are obtained by diagonalizing the Hamiltonian matrix,
where the superscripts $c$ and $v$ refer to the unoccupied conduction band and the occupied valence band, respectively.

When graphene ribbons and carbon nanotubes are under the influence of an electromagnetic field, the response to such a field is directly reflected in the optical absorption spectra. At zero temperature, only interband transitions obeying the selection rule $\Delta k_{y}=0$ occur due to the zero momentum of photons. Based on the Fermi Golden Rule, the spectral absorption function is
given by
\begin{eqnarray}
A(\omega )&\propto&\sum_{n^{v},n^{c}}^{}\int_{1stBZ}\frac{dk_{y}}{{
2\pi }}\left\vert \left\langle \Psi ^{c}_{k_{y}}(n^{c})\left\vert \frac{
\widehat{\mathbf{E}}\cdot \mathbf{P}}{m_{e}}\right\vert \Psi ^{v}_{k_{y}}(n^{v})\right\rangle \right\vert^{2} \\ \nonumber%
&\times& Im\left\{\frac{{
f[E^{c}_{k_{y}}(n^{c})]-f[E^{v}_{k_{y}}(n^{v})]}} {E^{c}_{k_{y}}(n^{c})-E^{v}_{k_{y}}(n^{v}){-\omega }{-\imath \Gamma }}
\right\},
\end{eqnarray}
where $\mathbf{P}$ is the momentum operator, $f[E_{k_{y}}(n)]$ is the
Fermi-Dirac distribution, $m_{e}$ is the electron mass and $\Gamma (=0.0015$ eV) is the phenomenological broadening parameter. The first term in Eq. (3), $\left\langle \Psi
^{c}_{k_{y}} (n^{c})\left\vert \frac{ \widehat{\mathbf{E}}\cdot \mathbf{P}}{%
m_{e}}\right\vert\Psi ^{v}_{k_{y}}(n^{v})\right\rangle $, is the velocity matrix
element denoted by $M^{c,v}_{k_{y}}(n^{c},n^{v})$.
By choosing the electric polarization $\widehat{\mathbf{E}}$ along the longitudinal $y$-direction and substituting Eq. (1) into $M^{c,v}_{k_{y}}(n^{c},n^{v})$, the velocity matrix element becomes
\begin{eqnarray}
M^{c,v}_{k_{y}}(n^{c},n^{v})&=&\mathop \sum \limits_{m,n=1}^N [
{A_{m}^{c}}^{*}(n^{c}){B_{n}^{v}(n^{v})}+{B_{n}^{c}}^{*}(n^{c}){A_{m}^{v}(n^{v})}] \langle a_{mk_{y}}|\frac{p_{y}}{m_{e}}|b_{nk_{y}}\rangle \\ \nonumber%
&\simeq& \mathop \sum \limits_{m,n=1}^N [{A_{m}^{c}}^{*}(n^{c}){B_{n}^{v}(n^{v})}+{B_{n}^{c}}^{*}(n^{c}){A_{m}^{v}(n^{v})}]\frac{\partial}{\partial k_{y}}\langle a_{mk_{y}}|H|b_{nk_{y}}\rangle.
\end{eqnarray}
The gradient approximation is used in the evaluation of $M^{c,v}_{k_{y}}(n^{c},n^{v})$ [47,57,58]. According to Eq. (2), only the off-diagonal terms in the Hamiltonian, which are associated with the nearest-neighbor hopping integrals and dependent on $k_{y}$, account for the contribution to the velocity matrix element. Therefore, the spectral absorption function $A(\omega )$ is calculated by summation over all possible optical transitions.

\vskip 0.6 truecm
\par\noindent
{\bf 3. Results and discussion}
\vskip 0.3 truecm

\vskip 0.6 truecm
\par\noindent
{\emph{3.1. Magneto-electronic properties}}
\vskip 0.3 truecm

Zigzag ribbons and armchair carbon nanotubes exhibit rich magneto-electronic structures. At $B_{0}$=10 T, the low-energy band structure of the $N=400$ flat zigzag ribbon possesses several composite energy bands (black curves in Fig. 1(b)), in which each one consists of the dispersionless QLL and the parabolic band. The unoccupied conduction bands (occupied valence bands) away from $E_{F}=0$ are assigned band indices $n^{c}=0,1,2,...$ $(n^{v}=0,1,2,...)$. The Landau quantization occurs at $E^{n^{c,v}}\propto\pm\sqrt{n^{c,v}B_{0}}$, and the dispersionless wave-vector range of the QLLs decreases with the increment of $n^{c,v}$. At $E_{F}=0$, the partial flat bands of $n^{c,v}=0$ are attributed to the combination of edge states and Landau states [31].

In curved ribbons, the electronic properties are significantly affected by
the effectively non-uniform magnetic field. For $\theta=5\pi/6$ and $B_{0}$=10 T (red curves in Fig. 1(b)), the wave-vector range of the $n^{c,v}=0$ flat bands is slightly reduced, but the dispersionless energy remains the same. Nevertheless, the $n^{c,v}>0$ composite bands are converted into oscillating parabolic ones, each with three band-edge states. Such states would induce the vHs in the DOS and thus the absorption peaks. As for the lower-$n^{c,v}$ subbands, electronic states in the vicinity of the middle band-edge states ($k_{y}=2/3$) still belong to the QLL states, while those around the other two are drastically changed by the finite-size confinement effect. With the increment of $n^{c,v}$, the energy difference between the oscillating subband and the QLL is enlarged due to a reduction of the magnetic-quantization effect. This also leads to a slight decrease in the amplitude of the oscillating subbands. At even higher energy levels (not shown), the band structure exhibits monotonous parabolic bands. The above-mentioned results directly reflect the difficulty in aggregating electronic states in a curved ribbon. The described phenomena become more obvious with an increase of $\theta$. That is to say, the geometric curvature is the critical factor in determining the magneto-electronic properties.

At $\theta =2\pi $, the $N=400$ zigzag ribbon is reduced to a (200,200) armchair carbon nanotube [43-45]. As a result of the periodic boundary condition, the carbon nanotube shows an entirely different 1D band structure (Fig. 1(c)). The monotonous parabolic bands dominate the electronic properties, and the band structure is symmetric about $k_{y}=2/3$. Furthermore, the absence of the open boundary condition makes the partial flat bands at $E_{F}=0$ disappear. Without an external field, the armchair carbon nanotube owns two linear bands intersecting at $E_{F}=0$ and parabolic bands with double degeneracy. These energy bands can be specified by the quantized angular momenta $J$. However, a perpendicular magnetic field causes the mixing of states, converts the linear bands into parabolic ones, and breaks the energy degeneracy. Intersecting at $E_{F}=0$, the lowest conduction band and the highest valence band are labeled by the band indices $n^{c}=0$ and $n^{v}=0$, respectively. Furthermore, the pairs of conduction bands (valence bands) intersecting at $k_{y}=2/3$ are denoted $n^{c}=1,2....$ ($n^{v}=1,2....$), and sequenced in accordance with the energies. These electronic states are mixtures of different $J$s' states. It is also noted that $n^{c}=0$ and $n^{v}=0$ are the only subbands that show a small dispersionless region, i.e., QLLs. The higher QLLs gradually form due to the stronger $J$-coupling created by the increment of the diameter and/or the field strength. These important differences between carbon nanotubes and curved ribbons indicate that the boundary conditions play a crucial role in the electronic properties.

Reflecting the main characteristics of the band structure, the density of states (DOS) is defined as
\begin{equation}
D(\omega )=\underset{n^{c},n^{v}}{\sum}{\frac{2{I}_{y}}{2N}} \int
_{1stBZ}\frac{dk_{y}}{{ 2\pi }}\frac{{\Gamma }}{{\pi \lbrack
(E_{k_{y}}^{c,v}(n^{c,v})-\omega )}^{2}{+\Gamma }^{2}{ ]}}.
\end{equation}
Under a uniform magnetic field, there are two kinds of prominent peaks corresponding to the VHS appearances in the energy spectrum. One kind is symmetric delta-function-like peaks resulting from both QLLs and the partial flat bands at $E_{F}=0$. The other kind exhibits an asymmetric square-root form attributed to the band edge of the parabolic dispersion. For the DOS of the flat ribbon shown in Fig. 2(a), there are only several delta-function-like peaks, among which their intensities diminish with the increment of quantum number $n^{c,v}$. The peak structure, however, is significantly changed in the curved ribbon, e.g.,  $\theta =5\pi /6$ case in Fig. 2(b). Besides one delta-function-shaped peak apparently remains at $\omega=0$, any other peak splits into three asymmetric square-root ones with lower heights and frequencies. The peak splitting and shifting are derived from the evolution of band structure that each dispersionless QLL changes into the oscillating parabolic band with three band-edge states (Fig. 1(b)).
On the contrast, the carbon nanotube exhibits a rather different structure of the DOS than that of the curved ribbon, as shown in the Fig. 2(c), where all the other peaks, except for the peaks at $\omega =0$ and $\omega =\sim\pm0.1$ eV, are contributed by the monotonous parabolic bands and display the asymmetric square-root form due to the weaker magnetic quantization effects. The verification of the inter-band transitions around the VHSs, as indicated in this figure, is described in the following section.
\begin{figure}[htbp]
\center
\rotatebox{0} {\includegraphics[width=14cm]{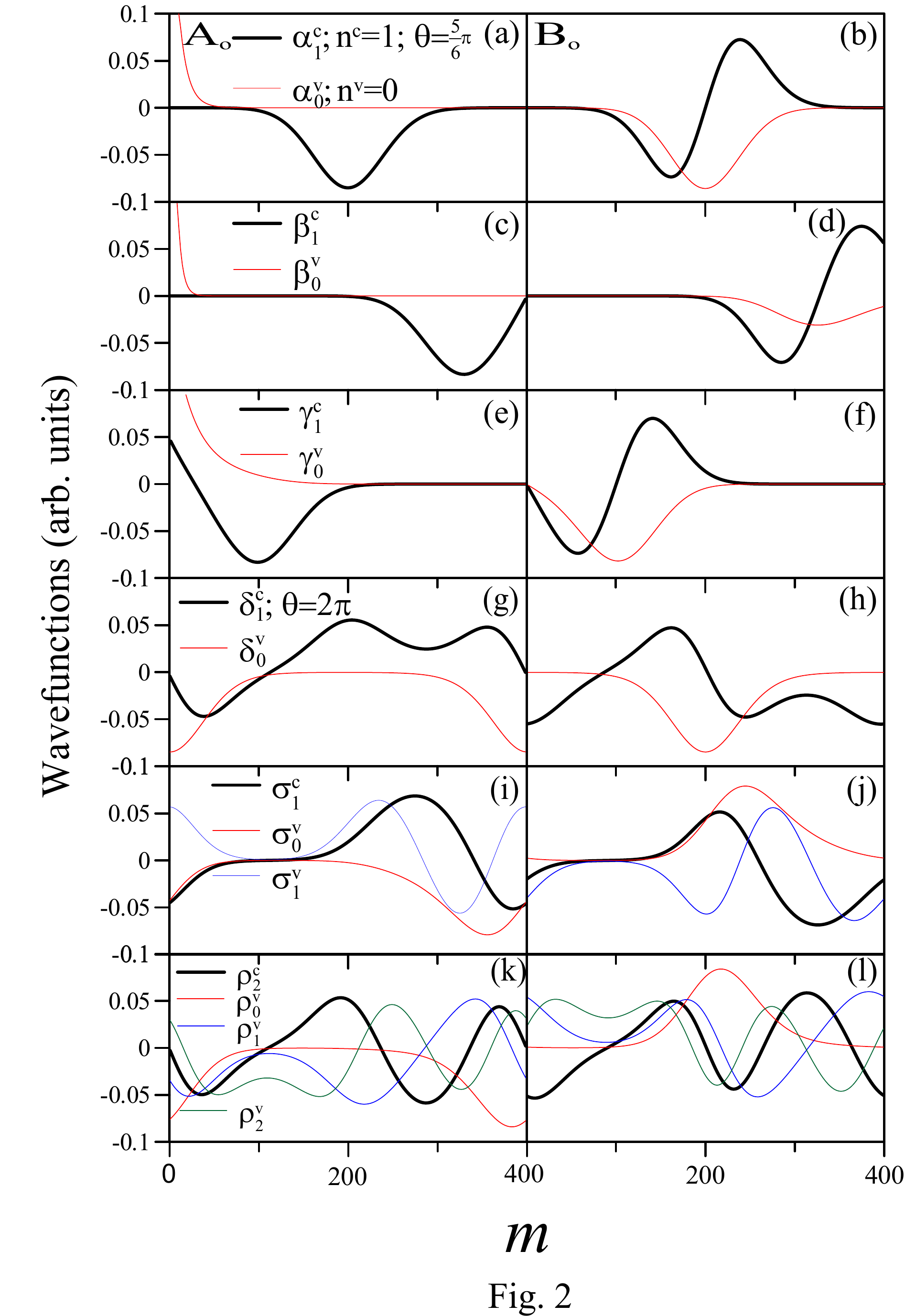}}
\caption{The density of states corresponding to (a) the $N=400$ flat zigzag ribbon, (b) the $N=400$ curved zigzag ribbon with $\theta $=$5\pi/6$ and (c) the (200, 200) armchair carbon nanotube under $B_{0}=10$ T. The absorption peaks resulting from the indicated transitions are presented in Figs. 4 and 5.}
\label{Figure 2}
\end{figure}

The spatial distribution of wavefunctions, which dominates the selection rules and absorption peaks, strongly depends on the geometric curvature and the field strength. For a zigzag ribbon or an armchair tube, the wavefunction can be decomposed into four components as follows:
\begin{eqnarray}
\left\vert \Psi _{k_{y}}^{c,v}(n^{c,v})\right\rangle
&=&\sum_{m,odd}^{N}\{A_{o}^{c,v}(n^{c,v},m)\left\vert \
a_{mk_{y}}\right\rangle +B_{o}^{c,v}(n^{c,v},m)\left\vert \
b_{mk_{y}}\right\rangle\} \\ \nonumber%
&+&\sum_{m,even}^{N}\{A_{e}^{c,v}(n^{c,v},m)\left\vert \
a_{mk_{y}}\right\rangle +B_{e}^{c,v}(n^{c,v},m)\left\vert \
b_{mk_{y}}\right\rangle \},
\end{eqnarray}
where odd- and even-related components are distinguished by the indices $o$
and $e$. The subenvelope function $A_{}^{c,v}(n^{c,v},m)$ ($%
B_{}^{c,v}(n^{c,v},m)$) represents the amplitude of the tight-binding function
on the $A_{}$ ($B_{}$) atom at the $m$th zigzag line. All the components are real numbers because its Hamiltonian elements, as expressed in Eq. (2), are all real. With the substitution of Eq. (2) and Eq. (6) into Eq. (4), the velocity matrix element becomes
\begin{eqnarray}
M^{c,v}_{k_{y}}(n^{c},n^{v})&=&\{\mathop \sum \limits_{m,odd}^N [
{A_{o}^{c}}(n^{c},m){B_{o}^{v}(n^{v},m)}+
{B_{o}^{c}}(n^{c},m){A_{o}^{v}}(n^{v},m) ] \\ \nonumber%
&+&\sum\limits_{m,even}[ {A_{e}^{c}}(n^{c},m){B_{e}^{v}(n^{v},m)}+
{B_{e}^{c}}(n^{c},m){A_{e}^{v}(n^{v},m)}]\} \\ \nonumber%
&\times&-\sqrt{3}b\gamma_{2}\sin(%
\sqrt[]{3}bk_{y}/2+\Delta G_{\mathbf{R}}).
\end{eqnarray}%
The Peierls phase difference $\Delta G_{\mathbf{R}}$ varies slowly in the primitive
unit cell; therefore, the last sine term can be taken out of the summation, so
\begin{eqnarray}
M^{c,v}_{k_{y}}(n^{c},n^{v})&\propto&\mathop \sum \limits_{m,odd}^N [{A_{o}^{c}}(n^{c},m){B_{o}^{v}(n^{v},m)}+{B_{o}^{c}}(n^{c},m){A_{o}^{v}}(n^{v},m) ] \\ \nonumber%
&+&\sum\limits_{m,even}^N[ {A_{e}^{c}}(n^{c},m){B_{e}^{v}(n^{v},m)}+
{B_{e}^{c}}(n^{c},m){A_{e}^{v}(n^{v},m)}].
\end{eqnarray}%
It is thus appropriate to only discuss the two $A_{o}$ and $B_{o}$ components, since the even- and odd-indexed envelope functions are just in opposite signs.

In a flat ribbon, the subenvelope functions of the $n^{c,v}\geq1$ Landau states, $A_{o}^{c,v}(n^{c,v})$ and $B_{o}^{c,v}(n^{c,v})$, have $n^{c,v}-1$ and $n^{c,v}$ zero points, respectively. They have the relationship $A_{o}^{v}(n^{v})=-A_{o}^{c}(n^{c})$ and $B_{o}^{v}(n^{v})=B_{o}^{c}(n^{c})$ for $n^{c}=n^{v}$. Moreover, the subenvelop functions can be described by the harmonic oscillators $\phi (n)$'s, each corresponding to the product of the $n$th orthogonal Hermite polynomial and a Gaussian function.
Therefore, $A_{o}^{c,v}(n^{c,v})\propto \ \phi(n^{c,v}-1)$ and $%
B_{o}^{c,v}(n^{c,v})\propto \ \phi (n^{c,v})$. It should be noted that a simple linear relationship exists between two sublattices and plays a dominating role in the optical selection rules, i.e., $A_{o}^{v}(n^{v})\propto B_{o}^{c}(n^{c}=n^{v}-1)$ and $B_{o}^{v}(n^{v})\propto
A_{o}^{c}(n^{c}=n^{v}+1)$. Specifically, the $n^{c,v}=0$ partial flat bands are characterized by the combination of two different components: the exponential decay function ($A_{o}^{c,v}(n^{c,v}=0)$) and the Gaussian function ($B_{o}^{c,v}(n^{c,v}=0)\propto\phi(n^{c,v}=0)$). However, the former, localized around one of the ribbon edges, only leads to a slight overlap with $B_{o}^{c,v}(n^{c,v}\geq 1)(\propto\phi(n^{c,v}))$, while the latter possesses the same features as aforementioned. Therefore, the relationships in two sublattices account for the special selection rule $\Delta n=n^{c}-n^{v}=\pm 1$ for transitions between the valence and conduction QLLs (Fig. 4(a); [41]).

In Fig. 3, we plot the representative wavefunctions to validate the possible interband transitions associated with the peaks in the DOS (Fig. 2) through a closer look at Eq. (8), which essentially constitutes a velocity matrix that is used to reveal the transition probability.
\begin{figure}[htbp]
\center
\rotatebox{0} {\includegraphics[width=14cm]{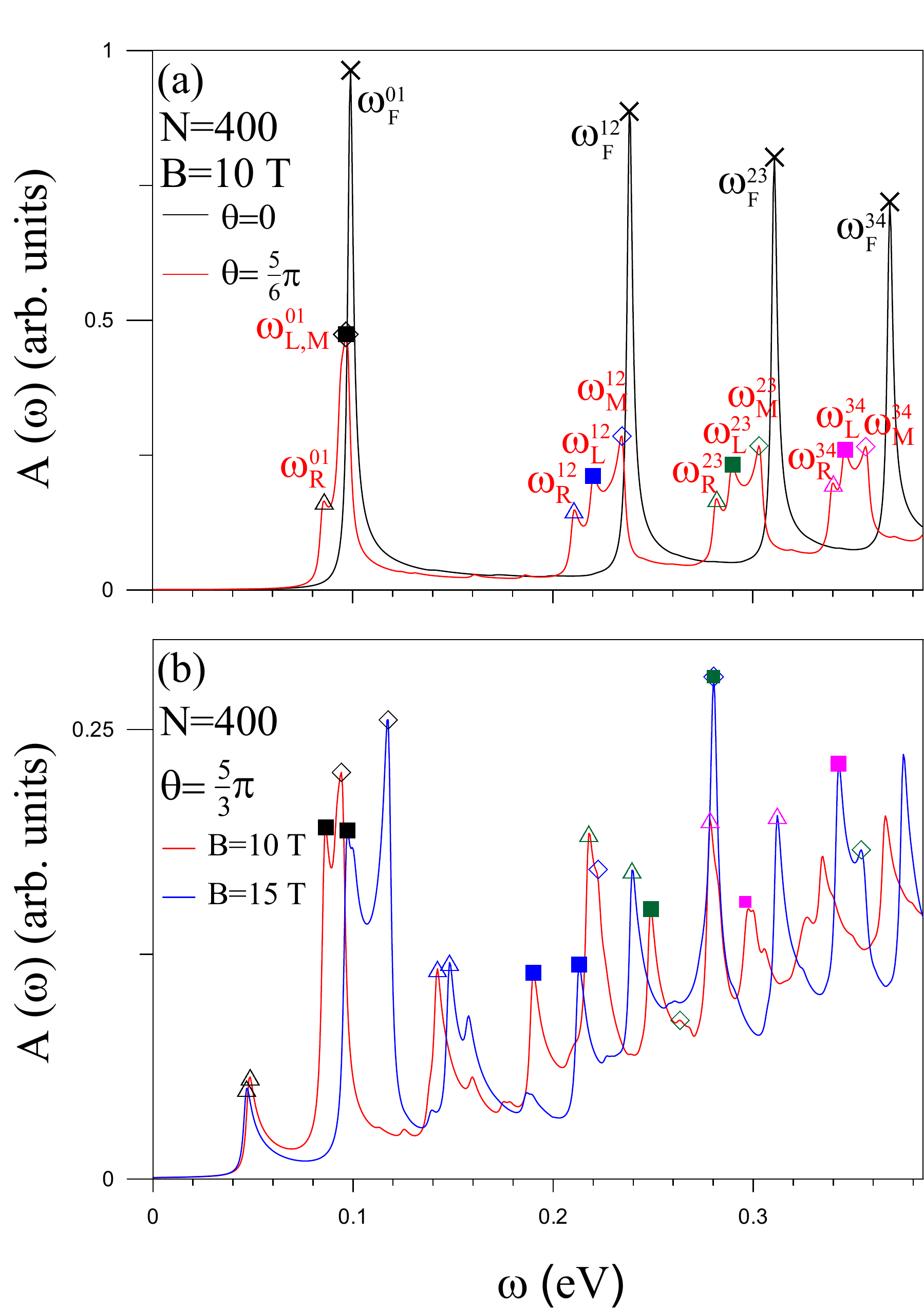}}
\caption{The odd-indexed wavefunctions corresponding to (a)-(f) states indicated in Fig. 1(b) of $\theta $=$5\pi/6$ curved graphene nanoribbon---the states of $\alpha _{0}^{v}$, $\alpha _{1}^{c}$, $\beta_{0}^{v}$, $\beta_{1}^{c}$, $\gamma_{0}^{v}$ and $\gamma_{1}^{c}$. For a carbon nanotube, (g)-(l) are the wavefunctions belonging to the states of $\sigma_{1}^{v}$, $\sigma_{1}^{c}$, $\delta_{0}^{v}$, $\delta_{1}^{c}$, $\rho_{0}^{v}$, and $\rho_{2}^{c}$ as indicated in Fig. 1(c).}
\label{Figure 3}
\end{figure}
The states $\alpha_{0}^{v}$, $\alpha_{1}^{c}$, $\beta_{0}^{v}$, $\beta_{1}^{c}$, $\gamma_{0}^{v}$ and $\gamma_{1}^{c}$, belonging to the $n^{v}=0$ or $n^{c}=1$ Landau subband, are illustrated in Figs. 3(a)-3(f).
For the $n^{v}=0$ states (red curves), $A_{o}^{v}(n^{v}=0)$ is present only near the edge, and the relationship between $B_{o}^{v}(n^{v}=0)$ and $A_{o}^{c}(n^{c}=1)$ is a determining factor during the optical excitations.
In Figs. 3(a) and 3(b), the simple proportional relationship between two subenvelope functions for the $\alpha _{0}^{v}$ state and the middle band-edge $\alpha _{1}^{c}$ state is presented, as born out by the fact that they still exhibit the spatial features of the Landau wavefunctions. However, for states deviating from the oscillating center $k_{y}=2/3$, the spatial symmetry of the Landau wavefunctions is broken. As to the $\beta_{0}^{v}$ state and the right band-edge $\beta_{1}^{c}$ state localized at one of the ribbon edges (Figs. 3(c) and 3(d)), the open boundary leads to a distortion of the wavefunction, while $A_{o}^{c}(n^{c}=1)$ and $B_{o}^{v}(n^{v}=0)$, overlapping strongly, have a nonzero velocity matrix element. The $\gamma_{0}^{v}$ state and the left band-edge $\gamma_{1}^{c}$ state exhibit a similar phenomenon, in which the wavefunction distortion occurs at the other geometric boundary (Figs. 3(e) and 3(f)). These results suggest that the optical excitations between the $n^{v}=0$ valence band and the $n^{c}=1$ conduction band are still allowed, just like those seen in a flat ribbon. In addition, the above results show that the states left-to and right-to the center point are affected differently owing to the magnitude of the effective field. This can explain the distinct energies of the three band-edge states in a subband, and thus the three distinguishable peak excitation energies. However, the distorted wavefunctions might induce extra excitation channels from the valence to conduction subbands. Considering the modification of higher inter-QLL transitions, much focus is put on the simple proportional relationship between the $A_{o}^{c,v}$ and $B_{o}^{c,v}$ subenvelope functions. Due to the wider spatial distributions for higher Landau states, the open boundary largely affects this relationship. A large $\theta$ enhances such modification, which can reduce the optical excitations of $\Delta n=\pm1$ and generate additional ones. The complexity of the absorption spectra is thus expected to increase with the curvature.

In a cylindrical carbon nanotube, the wavefunctions depend on the angular momenta $J$. Without a field, the subenvelope functions of quantum number $J$ are characterized by two linearly independent functions, $\sin (\pm J\Phi_{m})$ and $\cos (\pm J\Phi_{m})$, where $0\leq\Phi_{m}(=2\pi m/N)\leq 2\pi$; the $"\pm"$ signs specify double degenerate states. Given the orthogonality of trigonometric functions, the allowed optical transitions are only those satisfying the selection rule $\Delta J=0$ [47]. On the other hand, the coupled-state wavefunction in a perpendicular magnetic field can be expressed as a combination of different $J$'s. This gives rise to the available transitions between two states containing the same $J$ components. For example, as shown in Figs. 3(g)-3(l), $\delta_{0}^{v}$ consists of $J=0$ and $J=1$ components; $\sigma_{1}^{v}$ reveals the combination of $J=0$, $J=1$ and $J=2$ components; ditto for the other $J^{th}$-coupled states. Obviously, similar to the zero-field situation, the selection rule $\Delta n=0$ is also validated by many transitions, such as the one between states $\sigma_{1}^{v}$ and $\sigma_{1}^{c}$ (Figs. 3(g) and 3(h)). The additional transition between $\sigma_{0}^{v}$ and $\sigma_{1}^{c}$ is allowed following $\Delta n=\pm1$. The subenvelope functions of states $\delta_{0}^{v}$ and $\delta_{1}^{c}$, in Figs. 3(i) and 3(j), are mainly distributed around the locations of $m=1$ and $m=200$, where the penetration of the magnetic flux is at a maximum. This results in a significant wavefunction overlap, which makes transitions between $\delta_{0}^{v}$ and $\delta_{1}^{c}$ possible. Once the QLLs are formed, the spectrum exhibits more prominent peaks arising from channels obeying the selection rule $\Delta n=\pm1$. It should be noted that $\delta_{0}^{v}$ and $\delta_{1}^{c}$ are double degenerate states. Two pairs of interband transitions exist between these states. The figures only show one of the pairs, while the other, essentially equivalent, is not shown. Furthermore, in addition to $\Delta n=0$ and $\Delta n=\pm1$, the selection rule $\Delta n=\pm2$ is introduced in Figs. 3(k) and 3(l). It is thus deduced that when the field strength or tube diameter is increased, the stronger $J$-coupling induces further selection rules and absorption peaks.

\newpage
\vskip 0.6 truecm
\par\noindent
\emph{3.2. Curvature-dependent magneto-optical spectra}
\vskip 0.3 truecm

The above-mentioned magneto-electronic properties are directly reflected in many important features of the magneto-optical spectra. A flat ribbon possesses many prominent absorption peaks. In the case of $N=400$ and $B_{0}=10$ T, nonequally spaced peaks exist in the symmetric delta-function-like divergence form, as shown by the black curve in Fig. 4(a).
\begin{figure}[htbp]
\center
\rotatebox{0} {\includegraphics[width=13.5cm]{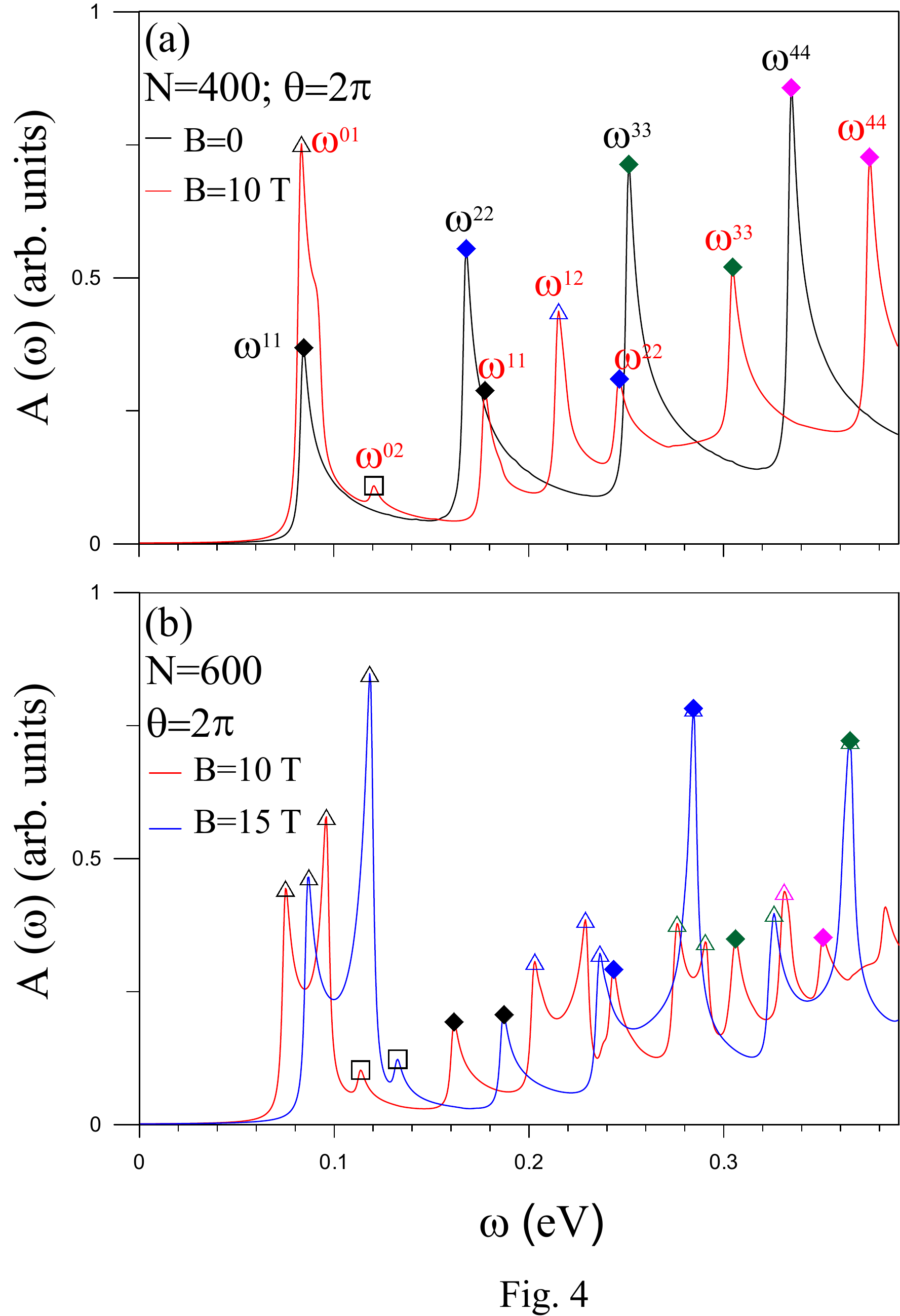}}
\caption{The optical absorption spectra of an $N=400$ zigzag graphene nanoribbon for (a) different arc angles ($\theta=0$ and $\theta=5\pi/6$) under $B_{0}=10$ T; for (b) $\theta=5\pi/3$ under different field strengths ($B_{0}$=10 T and 15 T). The three subpeaks of each group are denoted by open diamonds, solid squares and open triangles with the same color. The subscripts $R$, $L$ and $M$ are used to identify the transitions from the right, left and middle band-edge states of the Landau subband in Fig. 1(b).}
\label{Figure 4}
\end{figure}
Vertical transitions between valence and conduction QLLs that satisfy the selection rule $\Delta n=n^{c}-n^{v}=\pm1$ can produce such peaks. As a result of the band symmetry about $E_{F}=0$, the peak $\omega_{F} ^{n(n+1)}$ corresponds to two cases, of which one is derived from $n^{c}-n^{v}=1$ with $n^{v}=n$ and the other from $n^{c}-n^{v}=-1$ with $n^{v}=n+1$ ($n$ is an integer). The peak frequency $\omega_{F} ^{n(n+1)}$ is proportional to $\sqrt{n}+\sqrt{n+1}$ according to the energies of QLLs, as indicated by the transitions in Fig. 2(a), and the peak height is reduced with increasing frequency due to the smaller dispersionless wave-vector region of the QLLs.

The arc angle, however, significantly changes the features of magneto-optical spectra, such as the number, structures, heights and frequencies of the absorption peaks. The spectrum demonstrates more distinct peaks related to the vHs for $\theta=5\pi/6$, as shown by the red curve in Fig. 4(a). These peaks display the asymmetric square-root divergence form and their intensities are lower than those in the $\theta=0$ case. Such results, responsible for the 1D oscillating Landau subbands, are caused by the weakened magnetic quantization. Still, it is the selection rule $\Delta n=\pm1$ that dominates the magneto-optical spectrum when $\theta$ is below a moderate value, because the features of the Landau wavefunctions are retained. These peaks can be clearly classified into several groups characterized by the same interband transition. Each group consists of three subpeaks labeled $\omega _{R}^{n(n+1)}$, $\omega _{L}^{n(n+1)}$ and $\omega _{M}^{n(n+1)}$ in increasing order of the frequency (open diamond, solid square and open triangle in Fig. 4). The subscripts $R$, $L$ and $M$ are used to identify the interband transitions from the right, left and middle band-edge states of a Landau subband, as indicated in Fig. 2(b). When determining the peak heights, the velocity matrix is also a major factor, in addition to the DOS. Based on the wavefunction analysis, the subpeak $\omega _{M}^{n(n+1)}$, contributed by the inter-QLL transitions around $k_{y}=2/3$, is the highest, and the two $\omega_{L}^{n(n+1)}$ and $\omega _{R}^{n(n+1)}$ subpeaks are the next lower and the lowest, respectively, because they are significantly affected by the curvature and quantum confinements.
It should be noted that the combination of the subpeaks $\omega_{L}^{01}$ and $\omega _{M}^{01}$ boosts the intensity of the
second peak. The merged peak is close to the flat-case $\omega_{F}^{01}$, and its spacing with the neighboring $\omega _{R}^{01}$ is evaluated by the two excitation energies $\alpha_{0}^{v}\rightarrow\alpha_{1}^{c}$ ($\gamma_{0}^{v}\rightarrow\gamma_{1}^{c}$) and $\beta_{0}^{v}\rightarrow\beta_{1}^{c}$ as indicated in Fig. 1(b).
Nevertheless, the subpeaks of groups in the higher-frequency spectrum are clearly separated. No pronounced peaks result from the other optical selection rules of $\Delta n\neq\pm1$.

For a sufficiently large arc angle, optical transitions of $\Delta n\neq\pm1$ are greatly enhanced, while the opposite is true for transitions arising from $\Delta n=\pm1$. The main reason is that the spatial symmetry of the wavefunctions is severely broken, especially at higher energy. In contrast to the moderate $\theta$ case ( e.g., $\theta=5\pi/6$), the red curve in Fig. 4(b) shows much lower peak intensities for $\theta=5\pi/3$. Also, the red shift and the spacing are obviously more pronounced in each group. The $\Delta n=\pm1$ peaks remain sharp in the lower-frequency region. However, they are broadened with an increase of $\omega$, with some becoming more obscure, such as the peak $\omega_{M}^{23}$ ($\cong0.26$ eV; green open diamond), and others even vanish, such as $\omega_{M}^{34}$ (without red open diamond). This indicates that the $\theta$ influence on the magneto-optical properties is stronger at higher energies. Within the range of 0.1 eV$<\omega<$0.2 eV, some unmarked weak peaks originate from extra excitations of $\Delta n\neq\pm1$. The nature of the excited states reflects the significant changes on the distributions of the Landau wavefunctions. In addition to the curvature, the field strength also affects the spectral characteristics considerably. With the stronger magnetic quantization at $B_{0}=15$ T, the spectrum demonstrates more prominent structures where most peaks are blue-shifted and their intensities rise, as shown by the blue curve in Fig. 4(b). The peak $\omega_{M}^{23}$ ($\cong0.36$ eV; green open diamond), less pronounced at $B_{0}=10$ T, is greatly enhanced. Under different settings for the field strength and arc angle, the absence of specific selection rules for the peaks arising from $\Delta n\neq\pm1$ is mainly a result of the complex relationship between the geometric structure and the magnetic field.

Carbon nanotubes exhibit rather different selection rules than those of curved ribbons due to the cylindrical symmetry. As illustrated by the black curve in Fig. 5(a), the absorption spectrum for $B_{0}$=0 shows asymmetric peaks in the square-root divergence form.
\begin{figure}[htbp]
\center
\rotatebox{0} {\includegraphics[width=14cm]{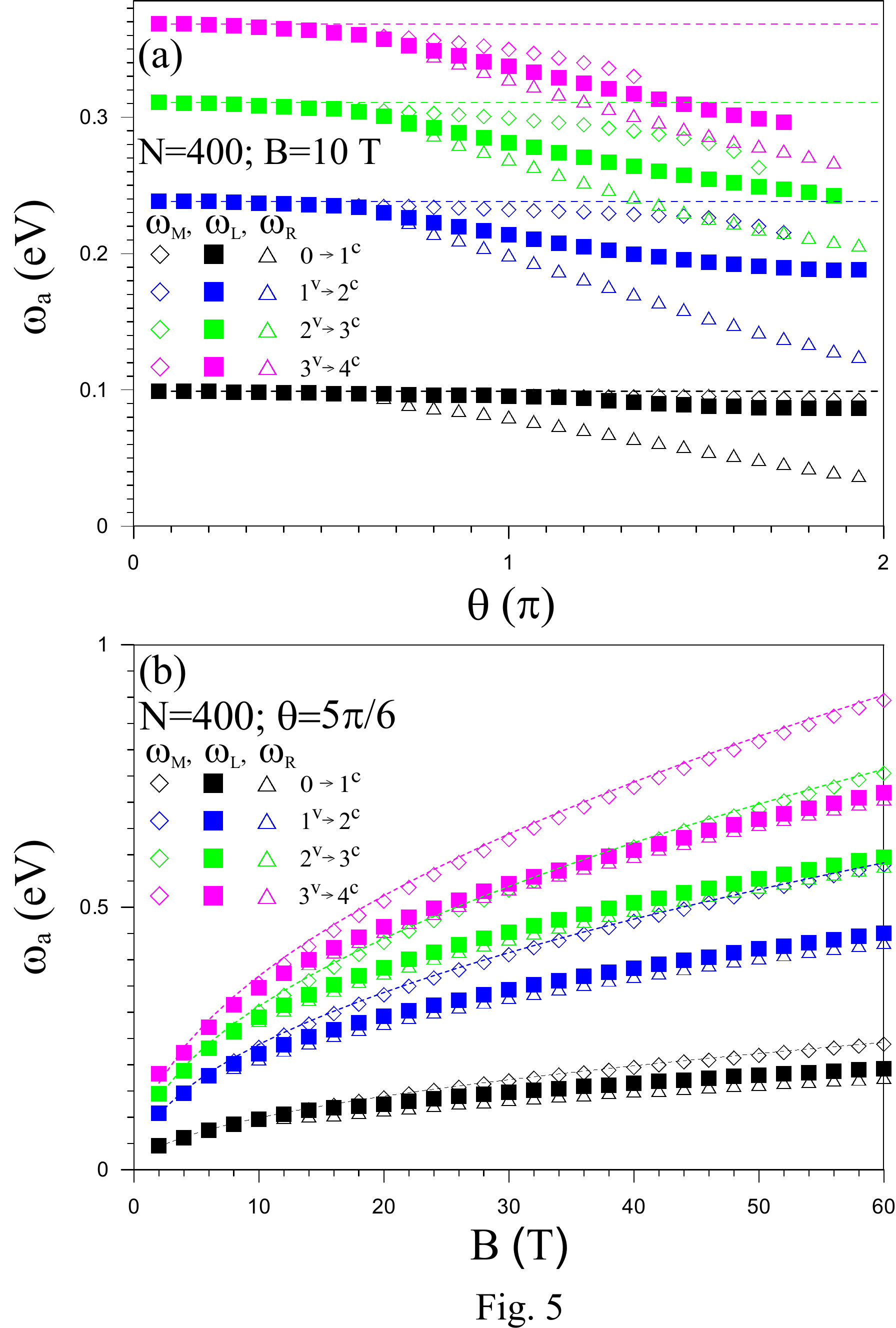}}
\caption{The optical absorption spectra for an armchair carbon nanotube with (a) $N=400$ at $B_{0}=0$ and $B_{0}=10$ T and (b) $N=600$ at $B_{0}=10$ T and $B_{0}=15$ T. The solid diamonds, open triangles and open squares denote the absorption peaks resulting from the selection rules $\Delta n=0$, $\Delta n=\pm1$ and $\Delta n=\pm2$, respectively.}
\label{Figure 5}
\end{figure}
Permitted by the selection rule $\Delta n=0$, each $\omega^{nn}$ is due to the transitions between the parabolic valence and conduction bands with the same band index $n$ (identical angular momentum). However, in the presence of a perpendicular magnetic field, besides $\Delta n=0$, other rules, such as $\Delta n=\pm1$ and $\Delta n=\pm2$, are introduced in the spectra through angular-momentum coupling. It is shown that at $B_{0}$=10 T (red curve in Fig. 5(a)), the original peaks $\omega^{nn}$'s are suppressed and shifted to higher frequencies. The extra peaks $\omega ^{01}$, $\omega ^{12}$ and $\omega ^{02}$, derived from $\Delta n=\pm1$ or $\Delta n=\pm2$, are, respectively, associated with the vertical transitions $%
\sigma_{0}^{v}\rightarrow\sigma_{1}^{c}$, $\varrho_{1}^{v}\rightarrow\varrho_{2}^{c}$ and $\varrho_{0}^{v}\rightarrow\varrho_{2}^{c}$ (Fig. 1(c)).
The first two $\Delta n=\pm1$ peaks are prominent, especially the $\omega ^{01}$ peak, while the third peak from $\Delta n=\pm2$ is less distinct.
This indicates that the $J$-coupling between two neighboring subbands is stronger. Additionally, the shoulder near the first peak $\omega ^{01}$ arises from the $\delta _{0}^{v}\rightarrow\delta _{1}^{c}$ transition at $k_{y}$=2/3, as indicated in Fig. 1(c). A prominent peak is formed once the low-energy QLLs form around $k_{y}$=2/3 (below).

At a fixed magnetic field, the $J$-coupling effects on the optical properties are more significant in a large nanotube. A rich absorption spectrum is presented for the $B_{0}$=10 T and $N$=600 case, as shown by the red curve in Fig. 5(b). It contains pairs of peaks induced by the selection rule $\Delta n=\pm1$, i.e., $\omega^{n(n+1)}$ and $\omega^{n(n+1)'}$ (n=0, 1, 2). Concerning the vertical transitions near $k_{y}=2/3$, the new peaks $\omega ^{01'}$, $\omega ^{12'}$ and $\omega ^{23'}$ are formed as a consequence of the formation of the low-lying QLLs. Evidently, the peaks $\omega ^{23}$ and $\omega ^{34}$ also arise from the strong coupling. At $B_{0}$=15 T, the three most noticeable peaks are mainly contributed by inter-QLL transitions, as shown by the blue curve in Fig. 5(b), since both the number and the dispersionless wave-vector range of the QLLs are increased by the enhanced magnetic quantization. Therefore, with an increase of diameter or field strength, the selection rule $\Delta n=\pm1$ gradually dominates the optical transitions as compared with $\Delta n=0$ and $\Delta n=\pm2$.

As evidenced by the aforementioned characteristics of the optical spectra, the curvature, field strength and boundary condition all play an important role for all curved ribbons and carbon nanotubes. In curved ribbons with an open boundary condition, the majority of transitions satisfy the selection rule $\Delta n=\pm 1$ and produce several peak groups, each consisting of three subpeaks, whereas a simple regularity is absent for the additional tiny peaks that appear under a larger curvature. In carbon nanotubes with a periodic boundary condition, however, the angular-momentum coupling allows excitations between various pairs of subbands. The spectrum shows regular peaks following the selection rules $\Delta n=0$, $\Delta n=\pm 1$, and $\Delta n=\pm2$. The specific peaks from inter-QLL transitions can be interpolated among the solutions obtained from graphene systems; however, many fascinating spectral features are demonstrated under the cooperative or competitive effects between the geometry and the magnetic field.

The frequencies of the principal absorption peaks, significantly affected by the
geometric structure and the magnetic field, deserve a closer investigation.
In Fig. 6(a), the $\theta $ dependence of the first four groups, $0^{v}\rightarrow 1^{c}$, $1^{v}\rightarrow 2^{c}$, $2^{v}\rightarrow 3^{c}$ and $3^{v}\rightarrow 4^{c}$, for the $N=400$ curved ribbon at $B_{0}$=10 T is presented.
\begin{figure}[htbp]
\center
\rotatebox{0} {\includegraphics[width=14cm]{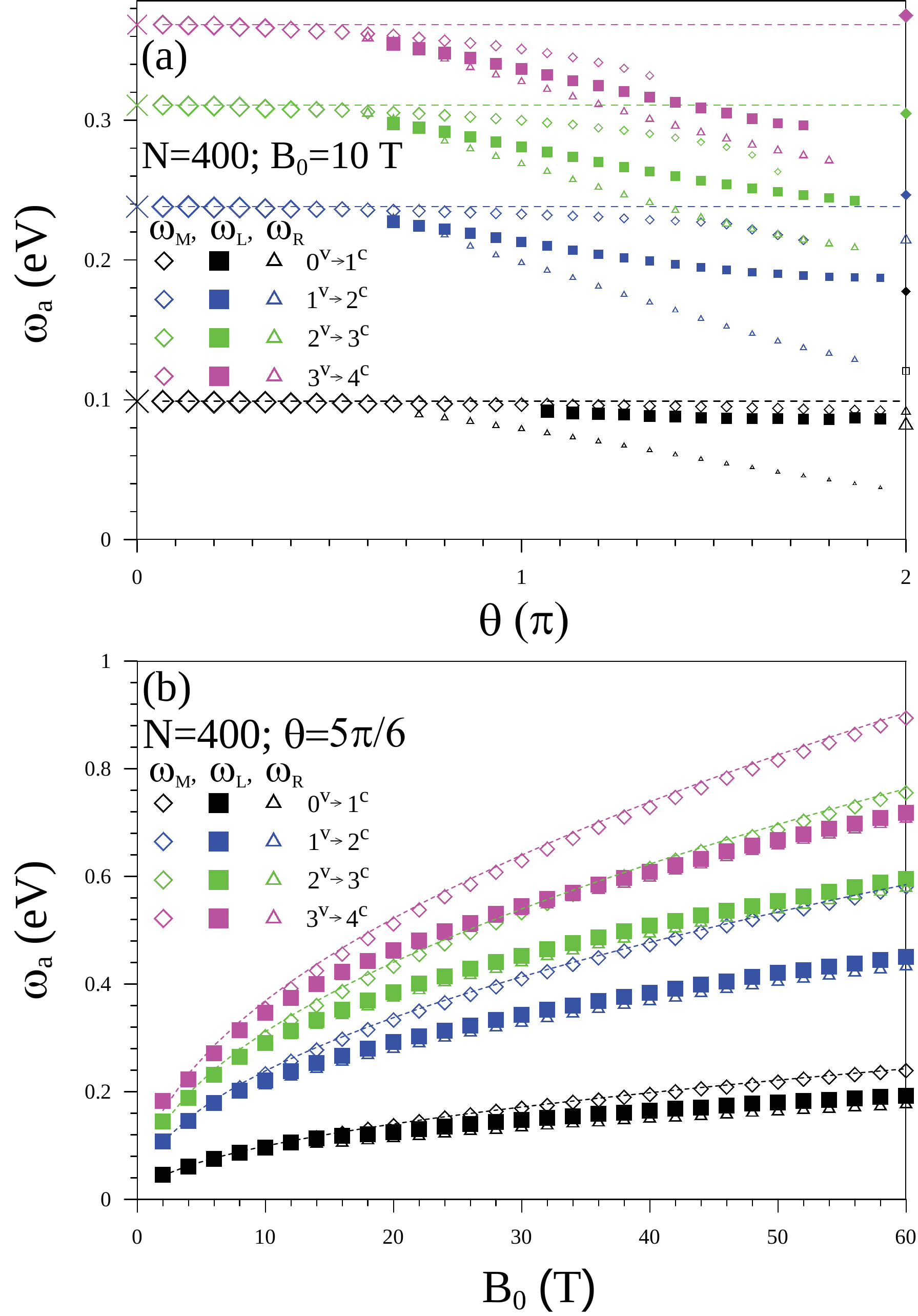}}
\caption{The $\theta$-dependent absorption frequencies for $N=400$ and $B_{0}=10$ T are shown in (a). The varying symbol sizes characterize the relative peak intensities, with the first inter-QLL peak in the flat ribbon chosen as the intensity unit. Regarding the $B_{0}$ dependence, the frequencies are plotted in (b) at $\theta=5\pi/6$. The dashed curves represent the inter-QLL transition frequencies.}
\label{Figure 6}
\end{figure}
With $\theta $ below 0.6$\pi $, the frequency of each group almost equals that of the flat-case peak (dashed line), which arises from the inter-QLL transitions. When $\theta $ further increases, three separate subfrequencies appear in each group as a result of the larger amplitude of the subband oscillation. Meanwhile, the red shift is strengthened by $\theta $; two subpeaks between adjacent groups, i.e., $\omega _{M}^{12}$ and $\omega _{R}^{23}$; $\omega _{M}^{23}$ and $\omega _{R}^{34}$, significantly overlap with each other for $\theta >1.4\pi $. Certain peaks even disappear when $\theta$ is above a critical angle because the QLL states are suppressed in the curved nanostructure. For the $n$th group, the critical angles of the three subpeaks are, in increasing order, $\theta _{M,C}^{n(n+1)}$, $\theta _{L,C}^{n(n+1)}$ and $\theta _{R,C}^{n(n+1)}$; that is, $\theta _{M,C}^{12}\simeq 1.73\pi $ for the second group, $\theta _{M,C}^{23}\simeq 1.67\pi $; $\theta _{L,C}^{23}\simeq 1.87\pi $ for the third group, and $\theta _{M,C}^{34}\simeq 1.33\pi $; $\theta _{L,C}^{34}\simeq 1.73\pi $; $\theta _{R,C}^{34}\simeq 1.87\pi $ for the fourth group . Figure 6(b) shows the relationship between the absorption frequency and the field strength for a curved ribbon with $N=400$ and $\theta =5\pi/6 $. The frequencies grow with the increment of $B_{0}$, but the curvature-induced splitting is absent until $B_{0}$ reaches 8 T. At high $B_{0}$, the three subpeaks in any group are clearly separated and the corresponding critical angles are expected to be enlarged. It should be noted that the intense peak $\omega _{M}^{n(n+1)}$ approaches a $\sqrt{B_{0}}$ dependence, which is a relationship identifiable by the inter-QLL transitions (dashed curves).

To understand the curvature effects on the nanotube spectra, the diameter dependence of the peak frequencies is investigated. At $B_{0}$=10 T, with various $N$'s from 100 to 700, Fig. 7(a) shows the frequencies of peaks coming from the interband transitions from the valence bands, $0^{v}$, $1^{v}, 2^{v}, 3^{v}$ and $4^{v}$, to the conduction bands, $0^{c}$, $1^{c}, 2^{c}, 3^{c}$ and $4^{c}$.
\begin{figure}[htbp]
\center
\rotatebox{0} {\includegraphics[width=14cm]{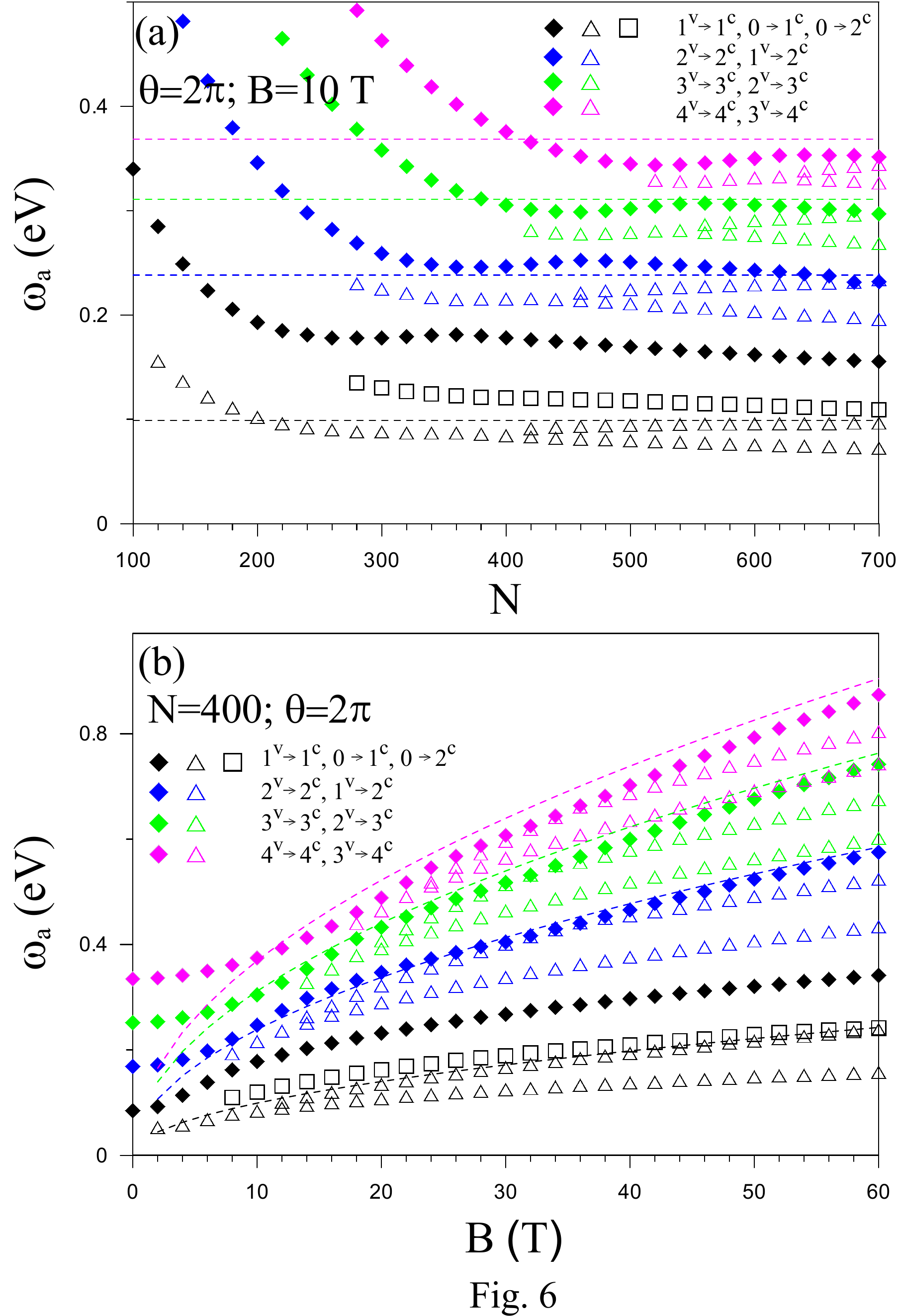}}
\caption{The dependence of absorption frequencies of carbon nanotubes on (a) the diameter at $B_{0}=10$ T and (b) the field strength for $N=400$. The inter-QLL transition frequencies are shown by the dashed curves.}
\end{figure}
The absorption peaks of $\Delta n=0$ at first quickly shift to lower frequencies with $N$ and then fluctuate within a narrow range. As a consequence of the stronger $J$-coupling, other peaks arising from $\Delta n=\pm1$ and $\Delta n=\pm2$ are gradually induced from the lower- to the higher-frequency region. The $\omega^{n(n+1)}$ peaks are the first to appear in each transition group ($0^{v}\rightarrow 1^{c}$, $1^{v}\rightarrow 2^{c}$, $2^{v}\rightarrow 3^{c}$ and $3^{v}\rightarrow 4^{c}$) and originate from excitations at both sides of the symmetric point $k_{y}=2/3$. When $N$ is above a critical value $N_{C}$, other peaks, $\omega^{n(n+1)'}$'s, occur next; their existence is attributed to the transitions between QLLs around $k_{y}=2/3$. The critical values $N_{C}$'s for the first four groups are $420$, $460$, $560$ and $640$. The increase of the estimated $N_{C}$'s can be explained by the fact that the higher QLLs only appear in large tubes. Similar to the stronger $J$-coupling with $N$, the stronger field also leads to more absorption peaks, as shown in Fig. 7(b). The critical field values $B_{C}$, which correspond to the onset of the QLL formation, are about 12 T, 14 T, 20 T and 24 T for the first four groups. It can be expected that such fields are relatively weak for a larger carbon nanotube, and that the square-root field dependence holds true for any of the groups in the case of $N>N_{C}$.

During the variation of $\theta$ from 0 to $2\pi$, the various peaks are exclusively formed in a flat or a curved structure with either open or closed boundaries. Regarding the different ribbon curvatures, the $\Delta n=\pm1$ peaks show splitting under the influence of a non-uniform effective magnetic field. The intensities and frequencies are continuously decreased with the increment of $\theta$. As the increment exceeds the critical values, some peaks vanish. When the complete continuity is interrupted for $\theta=2\pi$ (a tube), the selection rules caused by angular-momentum coupling, $\Delta n=0$, $\Delta n=\pm1$ and $\Delta n=\pm2$, account for the exclusive peaks of the nanotube.
The optical excitations under the respective boundary conditions, namely open and periodic ones, provide insight into the nature of the electronic properties in curved 1D systems. However, it is not until the diameters or field strengths exceed the critical values that the onset of peak pairs of $\Delta n=\pm1$ indicates the formation of QLLs. In particular, the square-root $B_{0}$-dependence of the absorption frequencies can evidently identify the QLL energy spectrum regardless of their nano-structures.
These theoretical predictions could be verified by means of spectroscopy experiments which provide the spectral features over a large energy range, including absorption peaks, spectral widths and intensities of different optical transitions [11,50-52,59].

\vskip 0.6 truecm
\par\noindent
{\bf 4. Conclusion}
\vskip 0.3 truecm

The low-frequency magneto-optical spectra of curved graphene nanoribbons and carbon nanotubes are studied using the Peierls tight-binding model. They are determined by the geometric curvature, boundary condition and magnetic field. The main characteristics of the absorption spectra, such as prominent structures, and number, intensity and frequency of absorption peaks, are thoroughly investigated.
The selection rules, deduced from the spatial symmetry of the wavefunctions, are rather different for these two systems due to the distinct boundary conditions.
The theoretical predictions can be examined by experimental optical measurements [11,50-52,59].

In the curved ribbons, the magneto-optical selection rule $\Delta n=\pm1$ remains dominant under a moderate variation of the arc angle since the wavefunctions retain the main characteristics of the Landau wavefunctions; however, the changes in absorption spectra are dramatic. Responsible for the low-lying oscillating and asymmetric Landau subbands, the absorption peaks show splitting and broadening with an increment of the arc angle, a response reflecting the effectively non-uniform magnetic field. These peaks are classified into several groups. Each group is comprised of three subpeaks, of which the highest one retains a strong magnetic quantization, while the other two are significantly influenced by the finite-size and curvature effects. At higher frequencies, the subpeaks become obscure and some even vanish when $\theta$ exceeds a critical value.
In addition, a small amount of relatively low peaks that appear irregularly with the controllable parameters are induced by the symmetry breaking of the Landau wavefunctions. The number and intensity of these peaks increase in curved ribbons with a larger curvature.

As a result of the angular-momentum coupling induced by a magnetic field, the optical selection rules in carbon nanotubes can continuously be altered by a change of the diameter or field strength. In addition to the selection rule $\Delta n=0$ in the absence of external fields, the strong coupling leads to other rules $\Delta n=\pm1$ and $\Delta n=\pm2$ that are allowed from the lower-frequency to the higher-frequency spectrum as a consequence of the increased diameter and/or field. The transitions following these rules produce a variety of regular peaks, in which the high intensity of the $\Delta n=\pm1$ peaks indicates the significant effect of the magnetic quantization. For a diameter (field strength) exceeding the critical value at a fixed magnetic field (diameter), the higher one in each pair of $\Delta n=\pm1$ peaks eventually matches the excitation energy triggered by the inter-QLL transitions. It is likewise demonstrated in the curved ribbons that even for a closed carbon nanotube, the intensities and frequencies of such peaks can be interpolated over the variation of curvature. A square-root $B_{0}$-dependence, an identifiable relationship of inter-QLL transitions, can also be observed, regardless of the geometric structure. However, the distinct spectral features of curved ribbons or tubes provide insight into the nature of the excitation states for the respective boundary conditions, namely open and closed (periodic)---features such as the specific selection rule, peak number, peak intensity and regularity of peaks.

Considering various chiral configurations of nanoribbons strained in the curved form, the low-energy bands under magnetic quantization will entirely evolve into oscillating Landau subbands. As for the optical properties with respect to the field strength and the geometric curvature, it is expected that the resulting magneto-optical absorption spectra show a strong resemblance. A prevailing phenomenon of carbon nanotubes is that the angular-momentum coupling induces extraordinary selection rules for any kind of chirality. The present work identifies the general effects of both the curvature and boundary on the optical responses of curved nanoribbons and carbon nanotubes under strong magnetic coupling.

\par\noindent {\bf Acknowledgments}

This work was supported in part by the National Science Council of Taiwan,
the Republic of China, under Grant Nos. NSC 98-2112-M-006-013-MY4 and NSC 99-2112-M-165-001-MY3.

\newpage
\renewcommand{\baselinestretch}{0.2}

\newpage \centerline {\Large \textbf {FIGURE CAPTIONS}}

\vskip0.5 truecm 

Fig. 1 - (a) A curved graphene nanoribbon, with the curvature radius $R$ and the bending arc angle $\theta$, in the uniform magnetic field $\textbf{B}=B_{0}\hat{z}$, perpendicular to the plane tangent to the ribbon bottom. At $B_{0}=10$ T, the band structures of (b) the $N=400$ curved zigzag ribbon with $\theta $=$5\pi/6$ and (c) the (200, 200) armchair carbon nanotube are plotted. Also shown in (b) is that of the flat zigzag ribbon for a comparison.

Fig. 2- The density of states corresponding to (a) the $N=400$ flat zigzag ribbon, (b) the $N=400$ curved zigzag ribbon with $\theta $=$5\pi/6$ and (c) the (200, 200) armchair carbon nanotube under $B_{0}=10$ T. The absorption peaks resulting from the indicated transitions are presented in Figs. 4 and 5.

Fig. 3 - The odd-indexed wavefunctions corresponding to (a)-(f) states indicated in Fig. 1(b) of $\theta $=$5\pi/6$ curved graphene nanoribbon---the states of $\alpha _{0}^{v}$, $\alpha _{1}^{c}$, $\beta_{0}^{v}$, $\beta_{1}^{c}$, $\gamma_{0}^{v}$ and $\gamma_{1}^{c}$. For a carbon nanotube, (g)-(l) are the wavefunctions belonging to the states of $\sigma_{1}^{v}$, $\sigma_{1}^{c}$, $\delta_{0}^{v}$, $\delta_{1}^{c}$, $\rho_{0}^{v}$, and $\rho_{2}^{c}$ as indicated in Fig. 1(c).

Fig. 4 - The optical absorption spectra of an $N=400$ zigzag graphene nanoribbon for (a) different arc angles ($\theta=0$ and $\theta=5\pi/6$) under $B_{0}=10$ T; for (b) $\theta=5\pi/3$ under different field strengths ($B_{0}$=10 T and 15 T). The three subpeaks of each group are denoted by open diamonds, solid squares and open triangles with the same color. The subscripts $R$, $L$ and $M$ are used to identify the transitions from the right, left and middle band-edge states of the Landau subband in Fig. 1(b).

Fig. 5 - The optical absorption spectra for an armchair carbon nanotube with (a) $N=400$ at $B_{0}=0$ and $B_{0}=10$ T and (b) $N=600$ at $B_{0}=10$ T and $B_{0}=15$ T. The solid diamonds, open triangles and open squares denote the absorption peaks resulting from the selection rules $\Delta n=0$, $\Delta n=\pm1$ and $\Delta n=\pm2$, respectively.

Fig. 6 - The $\theta$-dependent absorption frequencies for $N=400$ and $B_{0}=10$ T are shown in (a). The varying symbol sizes characterize the relative peak intensities, with the first inter-QLL peak in the flat ribbon chosen as the intensity unit. Regarding the $B_{0}$ dependence, the frequencies are plotted in (b) at $\theta=5\pi/6$. The dashed curves represent the inter-QLL transition frequencies.

Fig. 7 - The dependence of absorption frequencies of carbon nanotubes on (a) the diameter at $B_{0}=10$ T and (b) the field strength for $N=400$. The inter-QLL transition frequencies are shown by the dashed curves.

\end{document}